\documentclass[journal]{IEEEtran}
\usepackage{cite}
\usepackage{longtable}
\usepackage{url}
\usepackage[section]{placeins}
\usepackage{float}
\usepackage[cmex10]{amsmath}
\usepackage{array}
\usepackage{color}
\usepackage{epstopdf}
\usepackage{algorithm}
\usepackage{algorithmic}
\usepackage[pdftex]{graphicx}
\ifCLASSINFOpdf

\DeclareGraphicsExtensions{.pdf,.jpeg,.png}
\else
\fi
\begin{document}
%
\title {Machine Learning for Resource Management in Cellular and IoT Networks: Potentials, Current Solutions, and Open Challenges}
\author{
	Fatima Hussain,
	Syed Ali Hassan,
	Rasheed Hussain,
    Ekram Hossain
	
	\thanks{
		F. Hussain is with Royal Bank of Canada, Toronto, Canada (email: fatima.hussain@rbc.com).
		
			S. A. Hassan is with the School of Electrical Engineering and Computer Science (SEECS), National University of Sciences and Technology (NUST), Islamabad, Pakistan (email: ali.hassan@seecs.edu.pk).
		
		R. Hussain is with the Networks and Blockchain Laboratory, Institute of Information Security and Cyber-Physical System, Innopolis University, Innopolis, Russia (email: r.hussain@innopolis.ru).

		E. Hossain is with Department of Electrical and Computer Engineering at University of Manitoba, Winnipeg, Canada (email: Ekram.Hossain@umanitoba.ca). This work was supported in part by a Discovery Grant from the NSERC, Canada.
	}
}
\maketitle

\begin{abstract}
Internet-of-Things (IoT) refers to a massively heterogeneous network formed through smart devices connected to the Internet. In the wake of disruptive IoT with huge amount and variety of data, Machine Learning (ML) and Deep Learning (DL) mechanisms will play pivotal role to bring intelligence to the IoT networks. Among other aspects, ML and DL can play an essential role in addressing the challenges of resource management in large-scale IoT networks. In this article, we conduct a systematic and in-depth survey of the ML- and DL-based resource management mechanisms in cellular wireless and IoT networks. We start with the challenges of resource management in cellular IoT and low-power IoT networks, review the traditional resource management mechanisms for IoT networks, and motivate the use of ML and DL techniques for resource management in these networks. Then, we provide a comprehensive survey of the existing ML- and DL-based resource allocation techniques in wireless IoT networks and also techniques specifically designed for HetNets, MIMO and D2D communications, and NOMA networks. To this end, we also identify the future research directions in using ML and DL for resource allocation and management in IoT networks. 
 \end{abstract}

\IEEEpeerreviewmaketitle
\section{Introduction}
By 2020, it is expected to have 4 billion connected people, 25 billion embedded smart systems generating 50 trillion GB of data \cite{Fatima}. In the wake of such statistics, IoT, specifically wireless IoT,  has a great potential for the future smart world; however, the deployment of IoT on a massive scale also brings along lots of challenges. Some of these challenges include design of the networking and storage architecture for smart devices, efficient data communication protocols, proactive identification and protection of IoT devices and applications from malicious attacks, standardization of technologies, and devices and application interfaces. Additionally there are challenges such as device management, data management, diversity, and inter-operability \cite{Sen,Uday}.

For the wireless IoT, since radio resources are inherently scarce, resource management becomes very challenging in practical networks, specially when the number of contending users or smart devices is very large.  The overall performance of any wireless network depends on how efficiently and dynamically hyper-dimensional resources such as time slots, frequency bands and orthogonal codes are managed, and also how well the variable traffic loads and wireless channel fluctuations are incorporated into the network design in order to provide connectivity among users with diverse Quality-of-Service (QoS) requirements. Development of new IoT applications advocates for an improved spectral efficiency, high data rates and throughput as well as consideration of context and personalization of the IoT services, and consequently, the problem of resource management becomes crucial~\cite{Li,Li2018Access}.


\begin{table*}
\caption{Existing surveys on IoT}
\label{table:existingsurveys}
\begin{tabular}{|m{0.8cm}|m{0.8cm}|m{5.5cm}|m{2.3cm}|m{5.8cm}|}

\hline 
\textbf{Year} & \textbf{Paper} & \textbf{Topic(s) of the survey}  & \textbf{Related content in our paper}  & \textbf{Enhancements in this article}  \\ 
\hline
2015 & \cite{Fuqaha2015} & Enabling technologies, protocols and applications of IoT & Sec. \ref{sec:resourcemgtIoT} & Enhanced list of papers and coverage of more state-of-the-art solutions with focus on ML and DL \\
\hline
2017 & \cite{Gazis2017} & Communication standards in IoT & N/A & ML- and DL-based generic solutions in IoT\\
\hline
2016 & \cite{Ahmed2016} & Network types, technologies and their characteristics & Sec. \ref{sec:resourcemgtIoT} & Focus on state-of-the-art solutions in IoT and coverage of resource management, security, and privacy \\
\hline
2017 & \cite{Baker2017} & Health-care communications standards in IoT & N/A & Focus on generic IoT and in-depth coverage of the solutions \\
\hline
2018 & \cite{Javed2018} & Architecture, scheduling, network technologies, and power management of IoT operating systems & N/A & focus on the current solutions for generic IoT applications and future research directions \\
\hline
2018 & \cite{KOUICEM2018} & Blockchain and SDN solutions for IoT & N/A & Generic IoT and in-depth review of the ML- and DL-based solutions in multiple domains \\
\hline
2018 & \cite{CHOWDHURY2018} & Resource scheduling techniques in IoT & Sec. \ref{sec:resourcemgtIoT} & Detailed investigation of resource management in IoT and state-of-the-art techniques \\
\hline
2016 & \cite{AIREHROUR2016} & Secure routing protocols in IoT & N/A & Focus on the applications and resource management with security and privacy \\
\hline
2017 & \cite{GUO2017} & Trust models for service management in IoT & N/A & Coverage of overall aspects of security with solutions based on ML and DL \\
\hline 
2019 & \cite{DING2019} & data fusion in IoT applications through ML & N/A & In-depth coverage of ML and DL in different aspects of IoT \\
\hline
2018 & \cite{Mohammadi2018} & IoT data analytics through DL & N/A & In-depth review of ML- and DL-based solutions in IoT\\
 \hline
 \end{tabular}
\end{table*}

\subsection{Motivation of the Work}


\subsubsection{ Resource management issues in IoT networks} 


 \begin{itemize}
    
    \item {\em Massive channel access}:
   When a massive number of devices access the wireless channel simultaneously, it overloads the channel (e.g. Physical Random Access CHannel- PRACH in LTE networks). Also, a massive deployment of IoT devices bring challenges related to network congestion, networking and storage architecture for smart devices, and efficient data communication protocols  \cite{Sen,Uday}. Solutions for combating congestion challenges in access channel for IoT devices are mostly based on using different access probabilities and prioritization among different classes. Resource management for cellular IoT networks needs to consider load balancing and access management techniques to support massive capacity and connectivity while utilizing the network resources efficiently.
   
    \item {\em Power allocation and interference management}: In a dense (and large-scale) IoT network, intra- and inter-cell interference problems become crucial, and therefore, sophisticated power allocation and  interference management techniques are required. Many of the IoT applications can be of heterogeneous nature with potentially varying network topology, traffic as well as channel characteristics. Therefore, it is very challenging to choose transmit power dynamically in response to changing physical channel and network conditions.  
    
     \item {\em User association (or cell selection) and hand-off management}:
In cellular IoT applications, the devices have to associate with a Base Station (BS) or a gateway node to connect to the Internet. Again, the association may not be necessarily symmetric in both uplink and downlink scenarios (i.e. an IoT device may associate to different BSs at the uplink and the downlink). This association or cell selection affects the resource allocation. Also, for the mobile IoT devices,  resource management methods need to consider the possibilities of hand-off in different cells (e.g. macrocells, microcells).
 
\item {\em Harmonious coexistence of Human-to-Human (H2H) and IoT traffic}: 
For harmonious coexistence between the existing and new IoT traffic, intelligent resource management (e.g. for channel and power allocation/interference management, user/device association) is crucial that both the types of communications can co-exist with their requirements satisfied. 

\item {\em Coverage extension}: IoT devices are essentially low-power devices with limited computational and storage capabilities, in contrast to the traditional devices such as smart phones and tablets. This phenomenon reduces the coverage range. Therefore, to extend the coverage for IoT services, innovative resource management solutions are required. IoT devices can exploit Device-to-Device (D2D)  or relay-based communications techniques for their coverage extension. Dynamic resource allocation methods will be required for such communication scenarios.
 
 \item {\em Energy management}: Energy-efficient communication is very important  since most of the sensors and actuators in IoT  are battery-powered with limited battery capacity and charging facilities. Design of energy-efficient resource allocation and communication protocol design is essential to provide satisfactory QoS and support in dynamic network environments.
 
\item {\em Real-time processing and Ultra-Reliable and Low Latency Communication (URLLC)}:
 Mission-critical IoT applications such as remote surgery, public safety, Intelligent Transport System (ITS)  require low-latency communication and ultra-reliable connectivity. This calls for reliable and delay-aware resource allocation techniques. 
  
  \item {\em Heterogeneous QoS}: Since IoT networks are comprised of heterogeneous devices  with different  capabilities and  characteristics as well as communications requirements, they will require specialized and customized resources specific to their requirements. 
 
\end{itemize}

\subsubsection{Limitations of traditional resource management techniques}
Conventionally, the resource allocation problems are solved using optimization methods considering the instantaneous CSI and QoS requirements of the users. Generally the resource allocation problems are not convex and, therefore, the solutions obtained by using the traditional techniques are not globally optimal. Also, the solutions may not be obtained in real time. For practical implementation, more efficient (e.g. from computation as well as  performance point of view) solutions are needed. In some scenarios, where the resource allocation problem may not be well-defined mathematically (e.g. due to the nature of the propagation environment, mobility patterns of the users), we may not be able to formulate the optimization problem.  It motivates the exploration of new resource allocation schemes. In this context, the data-driven ML-based resource allocation solutions would be viable solutions in such scenarios and  should be adaptable to the dynamic nature of IoT networks.

\subsubsection{Why Machine Learning (ML) for resource management?}
ML can be defined as the ability to infer knowledge from data and subsequently to use the knowledge to adapt the behavior of an ML agent based on the acquired knowledge. ML techniques have been used in tasks such as classification, regression and density estimation. IoT devices generate sheer amount of data and the data-driven ML techniques can exploit these data to develop automated solutions for the IoT services. When the available data is massive and multi-dimensional, ML or more specifically, Deep Learning (DL) can be used for feature extraction and useful classification.

Also, we can use ML techniques when the prior knowledge of system, network, users, and parameters is not available and needs to be predicted and the control decisions need to be made. One such technique is Reinforcement Learning (RL) in which the unknown parameters and system behavior are monitored over time with {\it trial and error} and optimal control actions are learned. For resource management problems, when there is a {\em model deficit} or {\em algorithm deficit}, ML is recommended \cite{Simeone}. Model deficit means that there is insufficient domain knowledge or non-existent mathematical models, while algorithm deficit means that a well-established mathematical model is present but the optimization of the existing algorithms using these models is very complex. In such a situation, lower complexity solutions using ML are preferred. Also, when the contextual information is very important to be incorporated into the decision making process, ML techniques are most suitable. Most of the IoT applications fall under the above characterization because of the unknown system or network states and parameter values, and large number of devices generating enormous amount of data. 

To discuss the importance and requirements of ML techniques in various IoT applications, we focus on resource management in cellular IoT networks and smart home environment:

\begin{itemize}
\item {\em Resource allocation in cellular IoT networks}: Cellular IoT networks (e.g. 5G cellular and beyond)  should support extremely high data rates and require continuous connectivity to serve heterogeneous devices and users with varying QoS requirements. The transmitting devices are expected to autonomously connect to an Access Point (AP) or Base Station (BS) and access the frequency channels for achieving acceptable Signal-to-Interference Ratio (SIR). Furthermore, careful channel and power allocation is required to support various IoT applications simultaneously. For this purpose, obtaining the network parameters such as CSI, traffic characteristics, and demands of the users/devices are challenging, especially in case of serving highly mobile users. Also, estimation of the communication parameters such as the Angle of Arrival (AoA) and Angle of Departure (AoD) are important for millimeter wave communications. 

To address the above challenges, ML can be a suitable solution from both the network and user perspectives. IoT networks (network entities such as eNodeB)  are expected to learn the context and diverse characteristics for both human and machine users, respectively, in order to achieve optimal system configurations. Smart devices are expected to be capable of autonomously accessing the available spectrum with the aid of learning, and adaptively adjusting the transmission power for the sake of avoiding interference and for conserving the energy. For this purpose, Deep Reinforcement Learning (DRL) and linear regression can be used. In millimeter Wave (mmWave) communications, the values of CSI and AoA, which are not known a priori, can be estimated based on an RL technique, e.g. Q-learning. Location prediction of high-mobility users can be obtained  using Support Vector Machine (SVM), and Recurrent Neural Networks (RNNs) \cite{Qaio, Talieh}, and this location estimation can be used for resource allocation. 

Various correlated system parameters such as users’ position and velocity, propagation environment, and interference in the system can be taken into account to develop ML-based resource allocation methods. 
Furthermore, techniques such as Principal Component Analysis (PCA) and K-means clustering can be used for clustering various types of devices so that the available resources can be allocated according to the QoS requirements of these groups or clusters \cite{Quer,Jamal,Han,AlaganF1}. It is worth noting that traditionally, the optimization techniques cannot incorporate the context and are unable to react according to the changing environments \cite{Budinska,Moura}. Therefore, using these techniques may result in poor resource utilization. Also, the game theoretic approaches often do not incorporate heterogeneity of players as well as dynamic change in the network topology and number of players for resource allocation \cite{Moura}.

\item {\em ML for Resource management in smart home environment}:
Smart home applications are among the most popular IoT applications that involve a combination of heterogeneous and ubiquitous devices such as security cameras, hand-held scanners, tablets, smart appliance, and wireless sensors. All of these devices randomly access the network resources and have varying access and QoS  requirements. ML techniques such as Q-learning and multi-armed bandit can work well for scheduling and random access of the resources in the smart home environment. It is because these reinforcement learning methods learn and can adapt dynamically to the variation in the network environment \cite{Samad,Karim}. Also, $K$-means clustering and PCA can be used to group small sensors and aggregate the small payload data, respectively. Note that the traditional optimization and heuristics-based
techniques can be computationally very expensive to run on small, inexpensive, and energy-constrained sensors. Furthermore, traditional game theoretical approaches may not be suitable due to heterogeneity of the devices as well as 
overhead due to the information updates and exchange.
\end{itemize}

\begin{table*}
\caption{Acronyms and their descriptions}
\label{table:acronyms}
\begin{tabular}{|m{2.5cm}|m{5.5cm}|m{2.5cm}|m{5.5cm}|}
\hline 
\textbf{Acronym} & \textbf{Explanation}  & \textbf{Acronym}  & \textbf{Explanation}  \\ 
\hline
\hline
ML & Machine Learning & NOMA & Non-Orthogonal Multiple Access \\
 \hline
WSN & Wireless Sensor Networks & DL & Deep Learning \\
 \hline
 RL & Reinforce Learning & SVM & Support Vector Machine \\
 \hline
 PRACH & Physical Random
Access Channel & KNN & K-Nearest Neighbour \\
 \hline
 NB & Naive Bayes & NN & Neural Network \\
 \hline
 DNN & Deep Neural Network & CNN & Convolutional Neural Network \\
 \hline
 PCA & Principal Component Analysis & RNN & Recurrent Neural Network \\
 \hline
 MLP & Multi-Layer Perception & ELM & Extreme Learning Machine \\
 \hline
ITS  & Intelligent Transportation System & ANN & Artificial Neural Network \\
 \hline
 LSTM & Long-Short Term Memory & GAN & Generative Adversarial Network \\
 \hline
V2V  & Vehicle-to-Vehicle communication & DRL & Deep Reinforcement Learning \\
 \hline
 QoS & Quality of Service & CSI & Channel State Information \\
 \hline
 QoE & Quality of Experience & MIMO & Multiple-Input Multiple-Output \\
 \hline
 HetNet & Heterogeneous Network & D2D & Device-to-Device Communication \\
 \hline
 SDN & Software-Defined Network & CDSA & Control-Data plane Separation Architecture \\
 \hline
 RSSI & Received Signal Strength Indicator &  URLLC & Ultra-Reliable and Low Latency Communication \\
 \hline
 DDoS & Distributed Denial of Service & LPWAN & Low-Power
Wide Area Network \\
\hline
NB-IoT & Narrow Band Internet of Things & SINR & Signal-to-Interference-plus-Noise Ratio \\
\hline
BAN & Body Area Network & AML & Adversarial Machine Learning \\
 \hline
 NE & Nash Equilibrium & PSO & Particle Swarm
Optimization\\
 \hline
 GPU & Graphics Processing Unit & MCS & Modulation and Coding Scheme \\
 \hline
DOA & Direction-of-Arrival & D-SCMA & Deep learning aided Sparse Code
Multiple Access  \\
 \hline
 BER & Bit Error Rate & OFDM & Orthogonal Frequency
Division Multiplexing\\
 \hline
 DQN & Deep Q Networks &  SAX & Symbolic  Aggregate  approximation \\
 \hline
 MDP & Markov Decision Process & RBM & Restricted Boltzmann Machine \\
 \hline
 CSS & Cooperative Spectrum Sensing & CRN & Cognitive Radio Networks \\
 \hline
 TACL & Transfer Actor-Critic Learning & DPM & Dynamic Power Management\\
 \hline
 SMDP & Semi-Markov Decision Process & MWIS-DW & Maximum Weighted Independent Set problem with Dependent Weights\\
 \hline
 FIFO & First In First Out & RRM & Radio Resource Management\\
 \hline
 SIC & Successive Interference Cancellation & SDR & Software-Defined Radio \\
 \hline
 
 \end{tabular}
\end{table*}

\subsection{Scope of the Survey}

In the following, we succinctly summarize the published surveys and compare our survey paper with the existing surveys in terms novelty of covered topics and depth of contents. 

\subsubsection{Existing surveys}
 Table \ref{table:existingsurveys} lists the existing surveys related to IoT networks. Most of the published surveys do not focus on the ML and DL techniques in IoT networks. The current published surveys either focus on particular aspects of the IoT or do not cover the resource management in a holistic way. In Table \ref{table:existingsurveys}, we focus on the topics that are covered in the existing surveys and we reflect on the same topics (if applicable) in this article. Furthermore, we also categorically discuss the enhancements in this article. The surveys in Table \ref{table:existingsurveys} focus on enabling technologies, standards, architectures, scheduling, solutions, routing, data fusion, and analytics.  Although Artificial Intelligence (AI)-based and ML techniques have been used in IoT to address different challenges that were not addressed through traditional techniques, the role of ML and DL has not been covered in a holistic way in the existing surveys. 
 
 To fill in this gap, we investigate in-depth, the role of ML and DL in IoT networks and discuss the ML- and DL-based resource management solutions for wireless IoT. We have covered the latest surveys (till 2019) in this article. Note  that the role of ML and DL is multi-faceted in IoT, therefore covering the entire spectrum of applications of ML and DL techniques in not possible in one article. We only focus on the resource management aspect of wireless IoT networks.   

  \begin{figure*}
\centering
\includegraphics[width=1\textwidth]{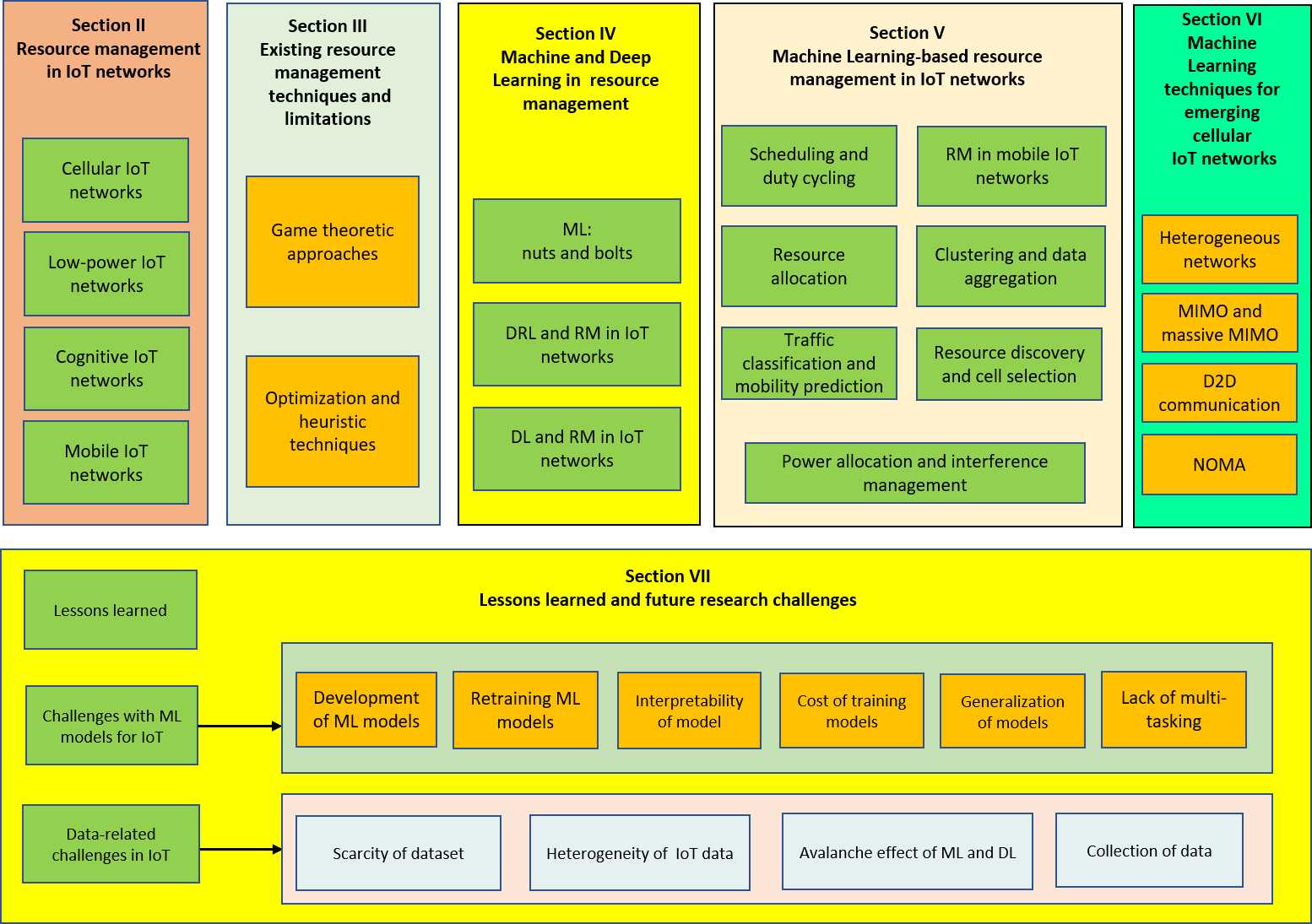}
\caption{Taxonomy of this`                                                   survey. [ML: Machine Learning, DL: Deep Learning, RM: Resource Management, DRL: Deep Reinforcement Learning, NOMA: Non-Orthogonal Multiple Access]}
\label{fig:taxonomy}
\end{figure*}

\subsubsection{Summary of the contributions}

We cover different aspects of resource allocation and management in IoT  networks that leverage ML and DL techniques. The objective of this survey is to bridge the gap between the resource allocation requirements of the IoT networks and the services provided by the ML and DL techniques.

Resource allocation and management are essential for the QoS guarantees in IoT applications and therefore need close attention. After covering the resource management challenges in different IoT networks (e.g. cellular IoT, low-power IoT, cognitive IoT, and mobile IoT networks), we discuss the existing solutions including those for Device-to-Device (D2D), HetNet, (massive) Multiple Input Multiple Output (MIMO), and Non-Orthogonal Multiple Access (NOMA)  systems. Then we motivate using ML and DL in addressing the resource allocation and management challenges in IoT that were not addressed by the traditional solutions. We then dive deeper into the ML and DL techniques that address the challenges of the traditional resource management and are leveraged for resource allocation in heterogeneous IoT networks. We also outline the future research opportunities and direction.  The pictorial illustration of the scope of this paper is shown in Fig. \ref{fig:taxonomy}. The contributions of this article are summarized as follows.

\begin{enumerate}
\item We discuss in detail the traditional techniques used for resource allocation in IoT network and their limitations that motivate the use of ML and DL techniques.

\item We conduct an in-depth and comprehensive survey on the role of ML and DL techniques in the IoT networks.

\item We discuss the existing solutions to IoT networks that leverage ML and DL algorithms with an emphasis on the major aspects of the resource management in IoT networks.

\item Based on the conducted survey of the ML- and DL-based resource management in IoT networks supporting D2D, MIMO and massive MIMO, HetNets, and NOMA technologies, we identify the existing challenges and future research directions. This will stimulate further research in  ML and DL for IoT networks.
\end{enumerate}

 The rest of the article is organized as follows: In Table \ref{table:acronyms}, we list all the acronyms used in the paper. The challenges related to resource management in IoT networks are discussed in Section \ref{sec:resourcemgtIoT}. In Section \ref{sec:resourcemglimitation}, we discuss the existing resource management techniques and their limitations followed by the role of ML and DL in resource management in Section \ref{sec:MLDLsolutions}. In Section \ref{sec:MLRMIoT}, we discuss the existing ML-based resource management solutions in IoT networks. Then 
 in Section \ref{sec:MLDLenabling}, 
 we review the ML and DL-based solutions for resource management for the emerging cellular wireless technologies for IoT networks. In Section \ref{sec:futurechallenges}, we discuss the future research challenges in developing ML- and DL-based solutions for resource management in IoT networks before we conclude the article in Section \ref{sec:conclusion}.

\section{Resource Management in IoT Networks}
\label{sec:resourcemgtIoT}

 In the following, we discuss the  resource management challenges in four different breeds of the IoT networks, i.e. cellular IoT networks, low-power IoT networks, cognitive IoT networks, and mobile IoT networks. 
 

\subsection{Cellular IoT Networks}
Cellular IoT networks can generally be characterized by the underlying technology they are using. For instance, the resource management techniques in different communication modes of a cellular IoT network (such as the Device-to-Device (D2D) mode,  HetNet mode, NOMA mode,  or dynamic spectrum access-based mode) are different than those in a homogeneous cellular mode of operation. 
Generally the issues related to resource management in D2D systems include session setup and management \cite{Doppler:2009:DCU:2288675.2294007}, channel allocation and interference  management in the underlay networks \cite{Wu:2015:PAD:2811961.2811990,5910123}. HetNets also suffer from the same problems of co-tier and cross-tier interference \cite{6824744}. To date, several studies have been conducted on HetNet models and their resource optimization \cite{National2015RadioRM,Tarapiah2015}.
Recently, NOMA has gained significant attention due to its potential to provide high spectral efficiency. Some papers have already published that focus on the resource management aspect and optimization of the network parameters \cite{7454317,7994888}. Traditional methods for resource allocation in cellular IoT networks rely mainly on the optimization techniques. These methods face challenges when the number of users become large and/or when the multi-cell scenarios are considered. This is because the optimization space becomes overwhelmingly large to cater for the entire network, and therefore, computational complexity of finding solutions increases significantly.

 \subsection{Low-Power IoT Networks}
Although the IoT networks offer both long-range and short-range communications through different technologies, there are many applications that require the energy portfolio of short-range technologies while demanding long-range communications. Therefore, low-power IoT networks especially those supporting long-range communications such as the Low-Power Wide Area Networks (LPWANs) become inevitable. LPWANs offer long-range communications by limiting the data rates and energy, and have been rigorously studied for various smart applications. Many technologies in the licensed bands (e.g. LTE-M, NB-IOT and EC-GSM) as well as those in un-licensed bands (e.g. LoRa, SigFox, Weightless) for low-power networks are under consideration which have to deal with different types of resource allocation problems \cite{7815384}. In most of the cases, the networks are characterized with a large number of nodes and the collisions among packets are avoided through scheduling and duty cycling \cite{MEKKI20191}. Similarly, the problems of interference in these networks arising due to the underlying modulation schemes result in performance degradation of the system \cite{8430542}. 
 
 \subsection{Cognitive IoT Networks}
 
 Cognitive networks have been an area of interest since last decade and many solutions have been proposed  to solve different resource management problems in cognitive networks. One of the salient features of a cognitive IoT network is that the nodes constantly search for resources opportunistically that are better suited for them to boost the overall performance of the network. The cognitive networks allocate resources of `primary user' to so-called `secondary user' when the primary user is absent or idle. However, as soon as the primary device is activated (the activity pattern of whom could be random), the secondary user has to vacate that channel. Therefore, resource allocation in these networks, which needs to consider primary users' activity and QoS requirements as well as secondary users' QoS requirements, wireless propagation, and other network parameters,  is very challenging \cite{5447050}. The channel sensing, detection, and acquisition are mainly done by static techniques that suffer from various imperfections such as collisions, increased outages, and decreased system throughput \cite{Nie2006}.     
 
 \subsection{Mobile IoT Networks}
 
 Mobile IoT (MIoT) is the extension of traditional IoT with mobility  \cite{Alnahdi2017}. More precisely, in MIoT, the IoT services and applications of IoT are mobile and can physically be transferred from one location to another. Another orthogonal variant of such IoT network is the Internet of Mobile Things (IoMT) where mobile $things$ are employed to form an IoT network \cite{Nahrstedt2016,Tcarenko2017}. Traditional IoT enables the communication among static $things$ to realize different services and increase the connectivity among objects and the physical world, whereas in MIoT and IoMT, the communicating entities, i.e. smart things move and maintain their inter-connection and accessibility. Traditional IoT scenarios such as smart home mimic the static connectivity among objects, whereas mobile objects such as vehicles, mobile robots, wearable devices add the mobility dimension to the IoT. Traditional applications and services realized through IoMT and MIoT include, but not limited to, smart industry with mobile robots, smart transportation with vehicles, and so on. 
 
 As afore-mentioned, the defining feature of the IoMT and MIoT is the node mobility. 
 Since mobility incurs extra control information to be exchanged among the network management entities,
resource management in IoMT and MIoT is more challenging than the traditional static IoT networks. In addition to these challenges, application context is also an important parameter to consider for resource management.

 
\section{Existing Resource Management Techniques and Limitations}
\label{sec:resourcemglimitation}
In this section, we  discuss the existing resource management tools and techniques in IoT networks and the enabling technologies.


\subsection {Optimization and Heuristic Techniques}
 
To date, many solutions have been proposed for efficient resource allocation in IoT networks using optimization and heuristics-based techniques. For instance, the authors in \cite{Oueis} used genetic algorithm for load balancing in fog IoT networks. However, the complexity of the algorithm increases with increasing the number of user requests and network size.  In the same spirit, the authors in \cite{MinHyeop} proposed a  resource allocation approach  using  genetic algorithm. They transformed the resource allocation  problem into a degree-constrained minimum spanning tree problem and applied a genetic algorithm  to solve the problem in a ``near-optimal"' fashion. However, the approach does not scale well in a dynamic IoT environment and may not be implementable in a practical scenario. Similarly, the authors in \cite{Mudassir} used Particle Swarm Optimization (PSO) based meta-heuristic for the distribution of blocks of codes among the IoT devices. They developed multi-component applications capable of being deployed on various IoT nodes, edge, and cloud. However, it works well in static environment and lack its applicability in multi-objective and dynamic environments.

In general, the traditional optimization and heuristics-based techniques have the following characteristics. 

\begin {itemize}

\item {\bf Inaccuracy of the model-based resource allocation methods}: Numerous model-oriented methods have been developed in the literature for resource management (e.g. for  power allocation and interference management). However, the mathematical models used for these methods, although analytically tractable, are not always accurate due to the  practical channel environment and hardware imperfections. In this context,
model-free and data-driven methods are potentially more promising. 

\item {\bf Cost and complexity:} Heuristic techniques (e.g. branch and bound) are also computationally expensive and incur significant timing overhead. Therefore, it will not suit well in delay-sensitive and real-time IoT applications.  Moreover, the computational complexity of these heuristics proportionally increases with the increase in the network size. Also, with evolutionary algorithms such as genetic algorithms, designing the objective functions and describing the parameters and the operators  can be difficult. 


\item  {\bf Convergence to local optima:} For sufficiently complex problems (e.g. non-convex problems), the task of obtaining a global optimum solution in the presence of many local optimum points is very challenging. 

\item {\bf Sub-optimal solutions in high dimensional problems:} With traditional heuristics-based algorithms, the quality of the solutions can degrade  with the increase in dimensionality of the problem. Heterogeneity and dynamically changing characteristics of the IoT networks call for more intelligent,  self-adaptive, and auto-configurable techniques.

\item {\bf Parameter sensitivity and lack of flexibility:} The solutions obtained through heuristics-based approaches are sensitive to the corresponding chosen system parameters. If the key decision variables and the constraints change (e.g. due to new devices added to the network), a hard- or pre-coded heuristic technique may no longer remain feasible and  it may need to be reconfigured for producing more viable solutions. Also, the traditional heuristics algorithms do not utilize the contexts of the users/devices and the network, and therefore, are unable to react to a dynamically changing environment. 

 
\end{itemize}

\subsection{Game Theoretical Approaches}

Game theory is generally used for distributed resource allocation in wireless and IoT networks in the presence of resource competition or cooperation among nodes (e.g. non-cooperative and cooperative game theory techniques, respectively).  To date, many game-theoretic mechanisms have been employed to address the resource management issues in IoT networks \cite{Semasinghe2017,Kharrufa2018,Zhang2017,Zhou2016,Zhang2017RAGame}. In \cite{Zhou2016}, the authors proposed a non-cooperative game-based resource allocation mechanism for LTE-based IoT networks. The authors focused on reducing the inter-cell and intra-cell interference and proposed an efficient power allocation algorithm. 
In another work, Zhang et al. \cite{Zhang2017,Zhang2017RAGame} proposed an optimized resource allocation framework for fog-based IoT. Specifically, the authors employed a Stackelberg game among data service operators, subscribers, and the fog nodes. The problem of computing resource allocation was formulated as a game between the service operators and subscribers.
Similarly, Semasinghe et al. employed non-conventional game theoretic approaches for resource management in large-scale IoT networks \cite{Semasinghe2017}. The authors used evolutionary, mean field, minority games, and other similar approaches to solve the resource management problems in IoT networks. 
Kharrufa et al. \cite{Kharrufa2018} used a game theory-based approach to optimize resource allocation, where the mobile nodes compete for network resources and aim at reaching a Nash Equilibrium (NE) solution for resource management. 
In fact, most of the game-theoretic models aim at obtaining the NE solution for the resource allocation problem.


Despite their advantages, there are some limitations of the game-theoretic approaches for resource management. We discuss the limitations as follows. 

\begin {itemize}
\item  {\bf Homogeneity of players and availability of network information:} Generally, NE solutions are based on the  assumption that all the players are homogeneous, have equal/similar capabilities, and have complete network information. However, heterogeneous IoT devices have different computational capabilities, use different communication technologies, and exhibit different behaviors. Therefore, the conventional NE solutions may not be appropriate for a practical IoT environment. Also, developing game models and obtaining the corresponding solutions for a system with incomplete and/or imperfect information can be very challenging. 

\item {\bf Lack of scalability and slow convergence:} 
The number of players plays an important role in the applicability of game-theoretic approaches. When the number of players increases, with a traditional game model, obtaining the solution of the model becomes more complicated. Although specific game models such as the evolutionary games or mean-field games are suitable for modeling competition among large number of nodes (i.e. players) \cite{Semasinghe2017}, due to the poor convergence properties, the distributed solutions obtained using these approaches may not be feasible for practical systems. Therefore, in an IoT network with large number devices, use of game-theoretic approaches for resource allocation can be limited.

\item {\bf Information exchange overhead:} 
Implementation of game-theoretic approaches may require a significant amount of information exchange (which will lead to delay), and is not practical due to massive number of devices in the IoT networks. These devices have limited memory and energy, and cannot maintain the records of all the actions pertaining to the devices in the network.

\end{itemize}


\begin{table}
\caption{Existing surveys on resource management using conventional methods}
\label{table:surveys}
\begin{tabular}{|m{5.5cm}|m{2.3cm}|}

\hline 
  \textbf{Topic(s) of the survey}  & \textbf{Related content}    \\
  
\hline
Device-to-Device (D2D) Communications & \cite{Doppler:2009:DCU:2288675.2294007}, \cite{Wu:2015:PAD:2811961.2811990}, \cite{5910123}, \cite{Haenggi:2009:SGR:1649956.1649958}, 
\\
\hline
	Heterogeneous Networks & \cite{6824744}, \cite{National2015RadioRM}, \cite{Tarapiah2015}, \\
	\hline
	Vehicular Networks & \cite{Zheng2015}, \cite{DBLP:journals/corr/abs-1801-02679} \\
	\hline
  Multiple Input Multiple Output (MIMO) and Massive MIMO & \cite{7593225}, \cite{7091863}, \cite{7447666}
  \\
  \hline
 	Non-Orthogonal Multiple Access (NOMA) & \cite{7454317}, \cite{7994888} \\
 	\hline
 \end{tabular}
\end{table}

\section{Machine Learning  and Deep learning for Resource Management}
\label{sec:MLDLsolutions}

 In this section, we discuss the basics of ML, DL, and DRL\footnote{A more detailed description of the ML and DL techniques can be found in  \cite{Fatima2019}.} and their suitability for resource management in IoT networks. 
 

\subsection{Machine Learning (ML) and Deep Learning (DL) Basics} 

Broadly speaking, ML is divided into three categories: supervised, unsupervised and reinforcement learning techniques.

 \begin{table*}

\caption{Research problems in IoT networks and applied machine learning techniques}
\label{table:research objective}
\begin{center}
\begin{tabular}{|p{5cm}|p{7cm}|}
\hline
{\bf Research Objective/Problems} & {\bf Machine Learning Techniques (Surveyed References)}\\
  \hline
  \hline
  Clustering and Data Aggregation & \begin{itemize}
  \item K-Means Clustering \cite{AlaganF1},\cite{Han}, \cite{Javed}, \cite{gupta2016novel}
  \item Fuzzy C-Means Clustering \cite{kumari2017semi}
   \item Supervised Learning \cite{chung2017automated}, \cite{khan2014novel} 
   \item Deep Learning  \cite{Chaoyun},  \cite{park2016situation}
  \item Principal Component Analysis (PCA) \cite{Quer}, \cite{Jamal}
      
  \end{itemize}  \\
    \hline
Resource Allocation (Scheduling, Random Access) & \begin{itemize}
      \item Q-Learning \cite {AlaganF1}, \cite{8057766},  \cite{7917970}
      \item Multi-Armed Bandit \cite{Maghsudi}
      \item Reinforcement Learning \cite{Reinfor}, \cite{Mao}
      \item DNN \cite{8114493}
      \item Simulated Annealing  \cite{BOHEZ2015109}
      \item Bayes Classification  \cite{7457131}
       \end{itemize}  \\
    \hline
    Traffic Classification and Mobility Prediction & \begin{itemize}
      \item  Clustering \cite {Tatsuya}
      \item Deep Learning \cite{Teng, Hyekim}
     \item Recurrent Neural Network 
     \item Supervised Inductive Learning  \cite{Via}
     \item SVM \cite{Qaio, Talieh}
    
       \end{itemize}  \\
    \hline
     Power Allocation and Interference Management  & \begin{itemize}
      \item Deep Reinforcement Learning \cite{FanMeng,Kang,Ning}
      \item CART Decision Tree \cite{Fan}
      \item Linear Regression \cite{Per}
        \item $K$-Means Clustering\cite{LiY}
       \item Supervised Learning \cite{Simone}
       \item KNN \cite{Vignesh}
       \item Q-Learning \cite{Shree,Arman, Salwa}
       \item Deep Learning \cite{Kulin}
       \end{itemize}  \\
    \hline
 Resource Discovery, Cell and Channel Selection &\begin{itemize}
      \item Deep Learning \cite{5589224},  \cite{8604101}
      \item Supervised Learning \cite{m3}, \cite{m13}, \cite{m10}
      
      \item PCA \cite{Chunxiao}
      \item Bayesian Learning \cite{Wen}, \cite{m7}
      \item KNN \cite{Chunxiao}, \cite{m6}
      \item Reinforcement Learning \cite{AlaganF1} \cite{Yi}, \cite{m5}
    \item Support Vector Machine \cite{Chunxiao}, \cite{m4}, \cite{m8}
   
      \end{itemize}\\
  
  \hline

      
\end{tabular}
\end{center}
\end{table*}

\textit{Supervised learning}:
These techniques use models and labels known a priori,  and can estimate and predict unknown parameters. Support Vector Machine (SVM), naive Bayes classifier, Random Forest, and Decision Tree (DT) are most commonly used supervised learning algorithms used for  classification and modelling of the existing data sets. This family of algorithms is mainly used for channel estimation, localization, spectrum sensing problems.

\textit{Unsupervised learning}: 
These techniques are used with unlabeled data and utilize input data in an heuristic manner. Potential applications of these algorithms in IoT networks include cell clustering, user association and load balancing.  

\textit{Reinforcement Learning (RL)}: 
RL techniques do not require training data sets and an agent learns by interacting with the environment \cite{Walid}. The objective is to learn the {\em optimal policy} (i.e. optimal mapping between state and action) so that the long-term reward is maximized for the agent. Q-learning is one of the most popular RL techniques. RL techniques can be utilized by the IoT devices and sensors for decision making and inferring under unknown and dynamically changing network conditions. For instance, RL can be utilized during spectrum sharing for channel access in cognitive radio networks (in the presence of unknown channel availability) and in small cell area networks (with unknown resource conditions). It can also be utilized for BS association, specially with  unknown energy and load status of the BS.
However, RL techniques may suffer from slow convergence  rate (in terms of learning the optimal policy at a system state)  and curse of dimensionality, which can create challenges for the RL techniques in dynamic IoT networks.

\begin{figure}

\centering
\includegraphics[width=0.4\textwidth]{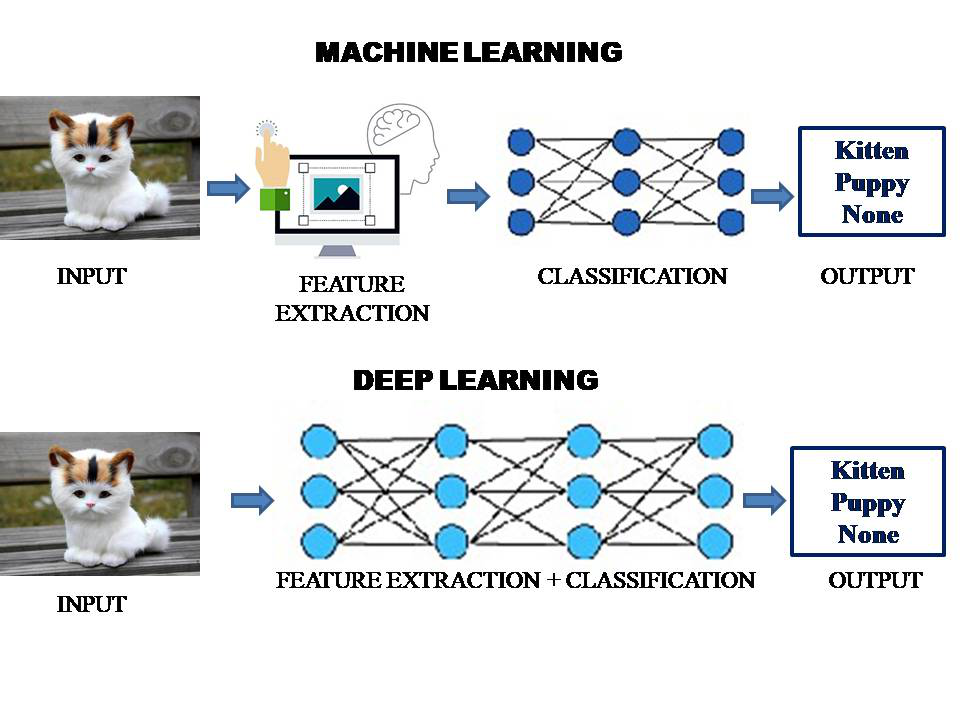}
\caption{Machine learning vs. Deep learning}
\label{deep}
\end{figure}

\textit{Deep Learning (DL)}: 
DL is a breed of the ML techniques which is based on Artificial Neural Networks (ANNs) with multiple hidden layers between the input and output layers. A Deep Neural Network (DNN), when trained, approximates the functional relationship that is present in a data set. It is expected that the relationship will be generalized for data points not present in the training data set.   DL is recommended for solving various problems in prediction, classification, resource allocation, spectral state estimation, object detection, speech recognition and language translation. DL learns features at multiple abstraction levels and allows a system to learn complex functions, and  maps the input to the output directly from data \cite{Wang}. For instance, in Fig. \ref{deep}, unlike ML, DL performs feature extraction and classification at various layers under the same process in which these are two separate processes. In DL, the first layer learns primitive features such as color and shape in an image by finding frequently occurring combinations of digitized pixels. These identified features are fed to the next layer, which trains itself to recognize more complex (combination of already identified) features such as edge, corner, nose, ears etc. of an image. These features are fed to the next layer which performs identification of more complex features (e.g. human, animal).  This recognition process is continued for successive layers until  the object is identified. The available open source DL frameworks/platforms include Torch/PyTorch, Thenano, TensorFlow and MXNet. DL is considered better than the traditional ML techniques due to following reasons: 

\begin{itemize}
 
 \item DL  is capable of handling large amounts of data and DL algorithms are scalable with increasing amount of data which can be beneficial to the model training and may improve prediction accuracy. 
 
  \item DL can automatically and hierarchically extract high-level features and corresponding correlations from the input data.  Therefore, the learning model is not comprised of multiple non-linear layers designed by the humans. It simplifies and automates the entire process. This automatic feature extraction, which is not very common in traditional ML techniques, can be beneficial for IoT networks where data generated by heterogeneous sources can exhibit non-trivial spatial/temporal patterns.
 
 \item IoT applications mostly generate unlabeled or semi-labeled
data. DL has the capability to exploit the unlabeled data to learn useful patterns in an unsupervised manner (e.g., using Restricted Boltzmann Machine (RBM) and  Generative Adversarial Network (GAN)). On the other hand, the conventional ML algorithms work effectively only when sufficient labeled data is available. 
 
 \item DL also reduces the computational and time complexity as single trained model can be used for various tasks and can achieve multiple objectives without model retraining.   Graphics Processing Unit (GPU)-based parallel computing enables DL to make predictions (such as resource allocations) very fast. Also,  when used along with technologies such as SDN and cloud computing, DL makes storage, analysis and  computation of data easier and faster. 
 \end{itemize}

However, the major drawback of DL is that a large amount of data will be required for training and testing, which may not be available or could be hard to generate. 

\textit{Deep Reinforcement Learning (DRL)}: 
DRL is a combination of DNN and RL where a DNN is used as an agent for RL. In other words, a DNN approximates the optimal policy function and it is trained using the data obtained through interaction with the environment. Therefore, no prior training data will be required. For problems in which the number of system states and data dimensionality are very large and the environment is non-stationary, using traditional RL to learn the optimal policy may not be efficient (e.g. in terms of convergence rate). By combining DNN and RL,  agents can learn by themselves and come up with a  policy to maximize long-term rewards. In this approach, the experience obtained through RL can help the DNN to find the best action in a given state \cite{Mufti, Mehdi,Ngoc,kai,Lake}. 

DRL is broadly divided into two categories: Deep Q Networks (DQN) and Policy Gradient. DQN is a value-based method while policy gradient is a policy-based method. In a 
value-based method, a $Q$ value is calculated for every action in the action space and selection of the action depends on the $Q$ value). In a policy-based method, the optimal policy is learned directly from a policy function (which gives the probability distribution of every action) without worrying about the value function (i.e. without calculating the Q-functions).  In DQN, Q-learning is combined with a DNN and the state action  value function ($Q$)  is determined by the DNN. In a policy gradient method, the policy with respect to the expected future reward is directly learned by optimizing the deep policy networks using gradient descent \footnote{https://medium.com/deep-math-machine-learning-ai/ch-13-deep-reinforcement-learning-deep-q-learning-and-policy-gradients-towards-agi-a2a0b611617e}. 


DRL is considered as a promising technique for future intelligent IoT  networks, and can be applied to  tasks with discrete and continuous state and action space, and also with multi-variable optimization problems.  DRL has been proposed for IoT networks for predictive  analysis and automation \cite{AlaganF1}. Detailed description of deep learning methodologies  
artificial neural networks can be found in \cite{Marco}. Authors provided comprehensive literature survey  on deep learning in wireless communication networks followed by various case-studies, in which DL has been useful. In the same spirit, authors presented and discussed various advance level deep Q learning (DQL) models  in \cite{Dusit}.  DQL  inherits  the advantages of DL and Q-learning and find its use fullness  in applications having large state and action spaces. To find the optimal policy and correlations between  the Q-values  and  the  target  value (with varying data  distribution), two  DQN models;  experience replay and target Q-network, are presented.

\subsection{Deep Learning and Resource Management in IoT Networks}

A flurry of recent research works has used DL techniques for solving the resource allocation  problems in IoT networks. For instance, a Q-learning based algorithm that uses cognitive radio-based IoT has been proposed in \cite{8057766}. The proposed algorithm uses DL to optimize the transmission of different packets of variable sizes through multiple channels so that the overall network efficiency is maximized. It has also been used in cognitive IoT networks  \cite{8604101,m2} where a cooperative strategy between secondary users is used to detect multiple occupancy of bands in a primary user.  In the same spirit, the authors in \cite{m12} proposed  DRL-based spectrum sensing approaches using which multiple pairs of secondary users coexist with these primary users. 


DL has also been used for advanced wireless transmission technologies such as MIMO and NOMA. For instance, \cite{huang2018deep} used DL for high-resolution channel estimation in MIMO systems. \cite{l8} proposed a DL technique, referred to as  DL-aided Sparse Code Multiple Access (D-SCMA) for code-based NOMA. An adaptive code book is constructed automatically which decreases the Bit Error Rate (BER) performance of a NOMA system. Using DNN-based encoder and decoder, a decoding strategy is learned and formulated. Simulation results for proposed scheme shows smaller BER and low computational time as compared to conventional schemes.

To this end, DL has also been used for load balancing  and interference mitigation in IoT networks. For instance, the authors in \cite{Hyekim} developed DL-based load prediction and load-balancing algorithms for IoT networks. Also, the authors in \cite{Kang} presented a DL-based interference minimization method.   Similarly, \cite{DBLP:journals/corr/OSheaEC17} used unsupervised DL in an autoencoder that optimizes the transmit parameters to provide diversity gains.
In \cite{8052521},  the authors provided a robust DL mechanism to estimate the CSI in an Orthogonal Frequency Division Multiplexing (OFDM) system, where the estimator performs better for pilot training and channel distortions. 


\subsection{DRL and Resource Management in IoT Networks}

DRL has been used for resource allocation problems to achieve high spectral efficiency. For example, the  authors in \cite{Li}  developed a deep Q-learning based spectrum sharing approach for primary and secondary users in which the primary users have fixed power control and the secondary users adjust their power after autonomously learning about the shared common spectrum. Similarly, the authors in \cite{Ekram-DL}, proposed a novel centralized DRL-based power allocation scheme for a multi-cell system.   Specifically, the authors used a Deep Q-learning (DQL) approach to achieve near-optimal power allocation and also to maximize  the overall network throughput.

Furthermore, in \cite{Challita}, the authors proposed a DRL-based  throughput maximization scheme in small cell networks. The DRL algorithm is developed by using a specific type of deep Recurrent Neural Network (RNN), called a Long Short-Time Memory (LSTM) network for channel selection, carrier aggregation, and fractional spectrum access.

In \cite{HaoYe},  a multi-agent DRL method was used for
resource allocation in vehicle-to-vehicle (V2V) communications. DRL is used to find the mapping between the local observations of each vehicle such as, local CSI and interference levels affecting the resource allocation decision. Each vehicle (agent) makes optimal decisions for transmission power by interacting with the environment.  
Ye et al. \cite{Ye2018resource} also used a DRL-based approach for resource allocation in vehicular networks. To reduce the transmission overhead, this approach uses local information on the available resources rather than waiting for the global information. The authors in \cite{Du2017} proposed a DL-based approach to predict the resource allocation based on 24 hour data count in V2V communication. He et al. \cite{He2018} proposed a DRL-based approach for optimized orchestration of computing resources in vehicular networks. Tan et al. \cite{Tan2018} used DRL in a multi-timescale framework for cache replacement and computing resource management. A detailed discussion about various DRL strategies can be found in \cite{Ahmed}.



\section{ML-Based Resource Management in IoT Networks}
\label{sec:MLRMIoT}
A summary of the ML-based resource management methods in IoT networks is provided in Table \ref{table:research objective}. A review of these methods is provided below.

\subsection{Scheduling and Duty Cycling}

For an energy-efficient operation, the nodes in an IoT network would be generally sending data intermittently and in short packets. In addition to traditional ways of controlling this duty-cycling, ML has been recently used to optimize these networks and system level parameters. In \cite{7457131}, the authors propose an ML-based up-link duty cycle period optimization algorithm for a broad class of IoT networks such as Long Range Wide Area Networks (LoRaWANs). More specifically, the authors optimize the up-link transmission period of LoRa devices using a Baye’s classifier of ML that improves the network energy efficiency. A scheduling method for an IoT network  for a smart home application was studied in \cite{7917970} where the authors compared various Q-learning algorithms for scheduling of home appliances.

Traffic scheduling in IoT systems need to consider various kinds of traffic such as delay-tolerant traffic and video and/or voice traffic. In \cite{Reinfor}, an RL-based traffic scheduling method was proposed that dynamically adapts to various traffic variations in a real-life scenario. More specifically, the authors applied RL algorithms on a 4-week real data traffic from a particular city and showed that the network performance almost doubles as compared to traditional scheduling. For a cloud-based IoT network, an ML-based task scheduling algorithm was proposed in \cite{8114493} which uses multiple criteria to optimize the network performance. 

\subsection{Resource Allocation}

ML can be used to solve the resource allocation problems in IoT networks  where a huge data set can be collected and used to train algorithms that produce very robust results for various resource allocation problems. In this context, \cite{8329631} uses cloud computing and ML together to perform resource allocation in a general wireless network and applies their proposed strategy for beam allocation in a wireless system.   A comprehensive analysis of cloud computing-based resource allocation algorithms was also discussed in \cite{8329631}.  Similarly, \cite{BOHEZ2015109} used different ML algorithms in a cloudlet-based network to efficiently optimize the network in terms of distributing computing and processing functions over the entire network entities. 

Note that, for the centralized wireless IoT networks such as cellular networks, resource allocation is generally performed by the BS based on the CSI of the users/devices and their Quality of Servie (QoS) requirements. However, for distributed IoT networks such as Machine-to-Machine (M2M) and ad-hoc networks, there is no centralized control and the users may not have the channel statistics available for the entire network. Channel allocation tasks in such scenarios can harness the benefit of state-of-the-art ML techniques.

ML-based resource management algorithms have also been studied for video applications in IoT networks. In \cite{6849102}, the authors use a mixture of supervised and unsupervised learning algorithms to optimize the Quality-of-Experience (QoE) of the end users when video streams are transmitted to them over a wireless channel. The learning algorithms extract the quality-rate characteristics of unknown video sequences and manage simultaneous transmission of the video over channels to guarantee a minimum QoE for users.  
Heterogeneous cloud radio access networks have also gained attention in the past few years for which the resource allocation and interference management are the key problems. Furthermore, in \cite{8292227}, the authors propose a centralized ML scheme to limit the interference below a certain threshold and it increases the energy efficiency of the network. A similar study on resource allocation in wireless systems is presented in \cite{Mao} where the authors use DRL algorithms to manage resources efficiently. 

\subsection{Power Allocation and Interference Management}

In \cite{Fan}, the authors used distributed Q-learning and CART Decision Tree algorithms for interference and power management in D2D communication networks. The authors were able to reduce the time complexity and improve system capacity and energy efficiency. In \cite{Per}, a system was proposed for interference level detection and power adaptation according to the interference level in the radio channels. The authors utilized a linear regression algorithm which uses CSI to predict the transmit power levels in order to minimize the interference and power wastage. Similarly, the authors in \cite{Ning}  proposed a hierarchical framework for resource allocation and power management in a cloud computing environment. The authors used a DRL method to develop a Dynamic Power Management (DPM) policy for minimizing power consumption.  

Collaborative distributed Q-learning was proposed in \cite{Shree} to reduce congestion in the access channel for resource-constrained MTC devices in IoT networks. The proposed system finds unique Random Access (RA) slots and reduce the possible collisions.  Similarly, in \cite{Vignesh}, the authors presented a model incorporating ML for low power transmission among IoT devices using LoRa\footnote{LoRa is a long-range low-power radio transmission technology which uses the license-free sub-GHz frequency bands.}. 
Moreover, the authors in \cite{Simone} developed a real-time burst-based interference detection and  identification model  for IoT networks. They developed an analytical model for interference classification by first extracting  Spectral Features (SFs) by exploiting the physical properties of the target interference signals. Afterwards, they used supervised learning classifiers such as multi-class SVM and Random Forest tree  for the isolation and identification of interference. 

\subsection{Clustering and Data Aggregation}
 \label{clustering}
 Clustering is one of the techniques to avoid overload and congestion in an IoT network. Clustering can be performed on the basis of parameters such as Euclidean distance, Signal-to-Interference-plus-Noise Ratio (SINR), device types, and QoS parameters. The most commonly used technique is $K$-means clustering algorithm, which is an unsupervised ML algorithm. 
$K$-means clustering partitions the available data set into $K$ clusters and every data point belongs to a cluster which has the least Euclidean distance from the centroid of that cluster. In \cite{gupta2016novel}, the authors propose a normalization algorithm to compute the distances iteratively, which increases the accuracy of clustering and also decreases the clustering time.

 The data streams generated in an IoT network can be huge and the data are generally unlabeled. The processing of raw  data is computationally expensive and manually labelling the data is not possible. In this context, \cite{khan2014novel} uses the Symbolic Aggregate approximation (SAX) algorithm to reduce the dimension of data stream and data is labeled using density-based clustering. The density-based clustering  extracts the number of classes in data and labels them using the found classes.


\subsection{Resource Discovery and Cell Selection}

A multi-band spectrum sensing policy was devised in \cite{5589224}, where the authors used an RL-based algorithm to sense the free channels of primary users. In \cite{m3}, the authors proposed a cooperative sensing scheme where multiple primary users with complex channel states cooperate using Extreme Machine Learning (ELM) algorithm to obtain accurate channel state  information and classification. This enables the secondary users to successfully access the idle channel which is not in use by primary user. Spectrum sensing can be very challenging in large-scale heterogeneous IoT networks. The authors in \cite{m7} proposed a new spectrum sensing scheme using Bayesian machine learning in which heterogeneous spectrum states are inferred using the Bayesian inference carried during the learning process.


In \cite{m4}, the authors used an SVM-based cooperative spectrum sensing technique. Similarly, in \cite{m5}, the authors presented an intelligent spectrum mobility management model for cognitive radio networks which can be used by the secondary users to switch among the spectrum bands. The authors used an ML scheme known as Transfer Actor-Critic Learning (TACL) was used for spectrum mobility management. For spectrum mobility, two modes are investigated: spectrum hand-off and stay-and-wait. In the spectrum hand-off mode, the user switches to another channel whereas in the stay-and-wait mode, the user stops transmission until the channel quality improves.  Also, long-term impact of the proposed scheme on network performance was analyzed in terms of throughput, delay, packet dropping rate, and packet error rate. Similarly, in \cite{m10}, the  authors proposed an online learning-based cell load approximation scheme in a wireless network.


\subsection{Traffic Classification and Mobility Prediction}  
  Traffic engineering with traffic classification as well as traffic (and hence mobility) prediction would be required  to accommodate the time-varying IoT traffic \cite{Tatsuya}. The authors in \cite{Via}  proposed a methodology  for construction of the model for normal behaviour for ongoing network traffic. Authors proposed traffic prediction procedure by separating short term (non-periodical or temporal)and longer-term(hour or day)variations. A longitudinal model  of network  connections was developed which can be used to identify deviations in the network traffic patterns. In \cite{Talieh}, the authors surveyed the flow-based statistical features of P2P traffic. They employed an SVM model to develop a system for classification of the encrypted BitTorrent traffic from usual Internet traffic.
  
  The authors in \cite{Yuan} proposed a learning-based network path planning model. They considered the sequence of nodes in a network path and determined implicit forwarding paths based on empirical data of the network traffic. The authors employed neural networks  for capturing  the data characteristics of the traffic forwarding and routing path. In \cite{Qaio}, the authors proposed a time window method  to acquire concise features from the packet header of network traffic for various applications. They used SVM, Back Propagation (BP) neural network, and Particle Swarm Optimization (PSO) to train and classify, and afterwards recognize the network traffic. In \cite{Parera}, the authors compared  the performance of six widely-used supervised ML algorithms (Naıve Bayes, Bayes Net, Random Forest, Decision Tree, and Multilayer Perceptron) for classifying network traffic.  The authors concluded that Random Forest and Decision Tree algorithms are the best classifiers for network traffic classifications, when classification accuracy and computational efficiency are required. 

\subsection{Resource Allocation in Mobile IoT Networks}
 Cellular networks, vehicular networks, mobile and ad hoc networks are examples of mobile IoT networks where the communicating devices are mobile. 
Various ML techniques have been proposed for modern cellular networks including D2D communication, massive MIMO and mmWave communications technologies as well as heterogeneous cellular networks (HetNets). {\em We will provide a survey of these techniques in the next section}. For Vehicular Ad-hoc NETworks (VANETs), the traditional approaches for resource allocation include dynamic programming \cite{Xing2014,Xu2015}, Semi-Markov Decision Process (SMDP) \cite{Li2017r}, Maximum Weighted Independent Set problem with Dependent Weights (MWIS-DW) \cite{Cao2016}, and users' aggregate utility maximization \cite{Qi2017}. Recently, ML algorithms have been used to address the resource optimization problem in vehicular networks. For instance, an ML algorithm was applied in \cite{6927124} to reduce the network overhead in a 
delay-tolerant routing system where repeated copies of a message signal are discarded.  


In the same context, over the last decade, vehicular networks have evolved to rich service and application space, for instance, vehicular clouds. In vehicular clouds, vehicles can form a cloud, can use the services of the cloud, or the combination of the two \cite{Hussain2015}. There are plethora of services realized through vehicular clouds such as traffic information as a service, cooperation as a service, storage as a service, and vehicle witness as a service \cite{HUSSAIN2015JPMC,Hussain2018,Hussain2014CooperationAwareVC}. In such scenarios, resource management is of paramount importance for efficient service provisioning. For this purpose, both ML and DL techniques have been proposed to allocate and manage resources in vehicular clouds. In \cite{Liang2018}, the authors explored the applications of ML and DL in managing the resources of vehicular networks. 

In \cite{Arkian2015}, Arkian et al. proposed a resource allocation mechanism for vehicular networks using cluster-based Q-learning approach. The authors aim at efficient cluster-head selection mechanism through employing fuzzy logic. The essence of the resource management scheme is to select the vehicle with enough resources to offer, in an efficient and QoS-aware way. In another work, Salahuddin et al. \cite{Salahuddin2016} proposed an RL-based technique for resource provisioning in vehicular clouds. The specific feature of the proposed RL-based resource management is that it provisions the resources based on a long-term benefit, which on one hand increases the utility, and on the other hand decreases the computational overhead.

\section{Machine Learning Techniques for Emerging Cellular IoT Networks}
\label{sec:MLDLenabling}

In this section, we survey the existing ML- and DL-based resource management solutions for specific cellular IoT technologies including HetNets, D2D, MIMO (massive MIMO), and NOMA. 

\subsection{ML Techniques in Heterogeneous Networks (HetNets)}
In a HetNet, cellular users coexist with small cell users using different Radio Access Technologies (RATs). The major issues related to resource management include mitigation of cross- and co-tier interference, mobility management, user association,  RAT selection, and  self organization.  In addition to the conventional methods (e.g. based on optimization or heuristics), ML-based methods have been used for resource management in HetNets \cite{8241779}. For instance, \cite{7127696} presented a mobility management method for a HetNet using RL, where inter-cell coordination and handovers are learnt by the devices to efficiently manage network resources. Similarly, \cite{7962716} reduced the energy consumption of the networks using a fuzzy game-theory approach where decision rules are learnt over time for better power consumption while maintaining a minimum QoS level for the users. The authors in \cite{m9} proposed a novel framework for cognitive RAT selection by the terminal devices in a HetNet using ML techniques.

 Recently, ML techniques have also been used to provide self-organization (which includes self-configuration, self-organization, and self-healing) in the networks. In \cite{Fan2014}, the authors studied an ANN approach to deal with self-optimization in HetNets. In \cite{8267327}, the authors proposed an efficient resource allocation algorithm for QoS provisioning in terms of high data rates that uses an online  learning algorithm and Q-value theory to optimize the resources. A similar work on resource allocation in  heterogeneous cognitive radio networks can be found in \cite{8292227}.      

\subsection{ML Techniques in Device-to-Device Communications}

In D2D networks, two devices in proximity connect to each other bypassing the central base station. By performing proximity communications, a D2D network offloads the traffic from the main BS and also improves the spectral efficiency as well as Energy Efficiency (EE) of the network \cite{8214255}. A low transmit power between the radios ensures EE, whereas low path-loss ensures high spectral efficiency and sum rate. There is a considerable amount of literature addressing various issues of D2D networks including resource and power allocation, mode selection, proximity detection, interference avoidance \cite{Liu2019,Ahmed2018}. Recently, ML has been used to address different problems related to D2D communications such as caching \cite{Cheng2019}, security, and privacy \cite{Haus2017} and so on. 

A major challenge in D2D networks is to efficiently utilize the limited resources to provide QoS requirements to all network entities, including cellular as well as D2D users. In \cite{D2D}, the authors devised a channel access strategy based on bandit learning in a distributed D2D system, where each pair selects an optimal channel for its communications. This does not only reduce the interference among users on the same channels, but also increases the rates of individual D2D pairs. Similarly, another work on channel selection using autonomous learning was presented in \cite{7517324}. A method for resource allocation in D2D networks was presented in \cite{7073432} where Q-learning is used to find the optimal resource selection strategy. Similarly, in \cite{fi9040072}, the authors used cooperative RL that improves the individual throughputs of the devices and the sum rates of the system using optimal resource allocation through a cooperative strategy planning. In \cite{7990595}, power allocations for various D2D pairs were optimally designed using ML to provide energy-efficient network solution.  

\subsection{ML in MIMO and Massive MIMO Communications}

Mobile nodes connected in a cellular system enjoy very high data rates if the number of antenna elements at the BS is large. Recently, a relatively new concept known  as massive MIMO has been an active area of research.  Many algorithms have been studied for channel estimation, beamforming, mitigating pilot contamination, and improving energy-efficiency in massive MIMO systems. 
In this context, ML techniques have also been used to address different problems in massive MIMO systems. For instance, in \cite{wen2015channel}, the authors addressed the channel estimation problem using sparse Bayesian learning algorithm. This algorithm needs good knowledge of the channel. Furthermore, the authors used the channel components in the beam domain as Gaussian Mixture (GM) distribution while Expectation-Maximization (EM) is used to learn the system parameters. The proposed channel estimator shows a good improvement as compared to the conventional methods in the presence of pilot contamination.

The positioning problem in MIMO systems was addressed in \cite{zhang2010mobile} using least square SVM. SVM finds the mapping between the sample space and coordinate space based on MIMO channel established by the multipath propagation. Simulations show improved location performance of the SVM as compared to ANN and  $K$-Nearest Neighbor (KNN). Furthermore, in \cite{jiang2017machine}, the authors recommended a supervised ML technique, i.e.,  SVM-based technique for MIMO channel learning. Similarly, in \cite{kim2018deep}, the authors proposed a supervised learning for pilot assignment to address the pilot contamination in MIMO systems, considering the output features as pilot assignment while the input features as user locations in the cells. This scheme is implemented  using a DNN. 

In \cite{dong2018machine}, the authors presented ML algorithms for link adaptation in massive MIMO. For the selection of transmission parameters, autoencoder-SVM and autoencoder-softmax were proposed. The autoencoder model is used to find features from CSI while logical regression models are used to choose modulation and channel coding algorithms. Simulation results show improved spectral efficiency compared to the conventional non-learning algorithms. Adaptive Modulation and Coding (AMC), which is also a part of link adaptation, is a challenging task in spatial and frequency selective channels. For the selection of suitable modulation scheme and coding rate in MIMO-OFDM systems, the past observations of the CSI and errors can be used. Similarly, in \cite{daniels2010adaptation}, the authors proposed a supervised learning method that uses the past information of the CSI and errors, to predict suitable modulation scheme and coding rate without finding the relation between input and output of wireless transceiver. To increase the accuracy of link adaptation using ML, low dimensional feature set using ordered SNRs was proposed. Simulation results showed that the proposed scheme improves the frame error rate and spectral efficiency. 


In another work, Mauricio et al. \cite{mauricio2018low} presented an ML approach to solve the grouping problem in multi-user massive MIMO systems. The mobile stations are classified into clusters using $K$-means clustering and then mobile stations are selected from the clusters to make an Space-Division Multiple Access (SDMA) group for capacity maximization. Similarly, in \cite{ 8644454}, the authors presented  a Radio Resource Management (RRM) and hybrid beamforming method for downlink massive MIMO system using ML techniques. The computationally complex resource management in multi-user massive MIMO is replaced with neural network using supervised learning. The results showed that the proposed ML-based method gives the same sum rate but decreases the execution time as compared to the performance of CVX-based method. Furthermore, Wang et al. \cite{ wang2018machine} proposed a ML-based beam allocation method for MIMO systems that outperforms the conventional methods.

\subsection{ML Techniques for NOMA}
For the next generation wireless networks, NOMA technique is a prime candidate to give spectral efficient performance and massive connectivity. The key idea behind NOMA is to handle multiple users in the same system resource unit \cite{l1}. These resource units are spreading code, time slot, sub-carrier, or space \cite{8472261}. Given the complexity of resource allocation problems in NOMA (i.e. user clustering, power allocation in the presence of imperfect Successive Interference Cancellation (SIC) and jamming attacks), it is of interest to exploit the benefits of learning-based technique  for NOMA systems \cite{l3}.

Notably, various works have recently applied ML and DL techniques for different  NOMA problems. For instance, in \cite{l7}, the authors proposed a novel approach using DL in NOMA system where single BS serves multiple users deployed in a single cell. The authors used a DL-based LSTM network which is integrated with traditional NOMA system. This enables the system to observe and characterize the channel characteristics automatically.
In \cite{l4}, the authors proposed an online learning based novel detection method for uplink NOMA scheme in a single cell. An online adaptive filter is designed for large cluster size of users, which provides better results than SIC-based detection method. In \cite{l5}, the authors proposed a MIMO-based anti-jamming game for NOMA transmission. They used a Q-learning-based allocation of power coefficients for transmission  in dynamic downlink NOMA  to counter  jamming attacks. This mechanism improves the NOMA system transmission efficiency in the presence of smart jammers. 

In \cite{l6}, the authors investigated the key problems of mmWave-based NOMA system, i.e., power allocation to the users and clustering of users. They derived an efficient user clustering for NOMA system. Considering continuous arrivals of users in the system, an online user clustering algorithm on the basis of $K$-means clustering was proposed which greatly reduces the computational complexity of the NOMA system. 
Similarly, in \cite{l9}, the authors proposed a NOMA-based resource management approach in a multi-cell network.  Furthermore, in \cite{l10}, the authors investigated the resource sharing problem for downlink transmission by considering maximum power constraint in NOMA networks. Moreover, in \cite{l11}, the authors dealt with  the resource allocation problem in a NOMA-based heterogeneous IoT network and proposed a deep RNN-based algorithm for optimal solution. The results showed that the proposed DL-based scheme outperforms the traditional schemes in terms of connectivity scale and spectral efficiency.

\section{Lessons Learned and Future Research Challenges}
   \label{sec:futurechallenges}


\subsection{Lessons Learned}

Resource allocation in IoT networks is a fundamental problem that becomes increasingly complex with the consideration of advanced wireless  technologies as well as system uncertainties (e.g., due to dynamic  channel and traffic variations, device mobility and multi-dimensional QoS requirements) and large-scale nature of the system (e.g. with thousands of devices and network nodes). Resource allocation problems are generally posed as optimization problems with one or more objective functions and some constraints. However, in many cases the conventional optimization problems are not  solvable with reasonable complexity, unless the problems are simplified and/or the constraints are relaxed. Also, when the problems are non-convex, they often provide only sub-optimal solutions. To reduce the complexity of obtaining resource allocation  solutions, new methods will be required. Also, in many cases the resource allocation problems cannot be modeled analytically due to the lack of appropriate modeling tools. 

Another common approach for resource allocation in distributed IoT networks is the use of game-theoretic approaches. However, modeling and analysis of the game models for large-scale  heterogeneous IoT networks under practical system constraints (e.g. uncertainty in terms of the number of devices and network characteristics, lack of availability of information about other devices) is extremely challenging. Also, even if the equilibrium solutions can be characterized, obtaining those solutions instantly would be a very highly ambitious goal due to slow convergence rate. 
  
In the above context, increasing interest in the interdisciplinary research has led to the use of ML, DL, RL, and related tools to make IoT devices and networks smarter and intelligent, and adaptive to their dynamic environments. Decrease in the computational costs, improved computing power, availability of unlimited solid-state storage, and integration of various technological breakthroughs have made this possible. Applications of ML and DL techniques enable IoT devices to perform complex sensing and recognition tasks and enable the realization of new applications and services considering the real-time interactions among humans, smart devices and the surrounding environment. DL, RL, and DRL algorithms can be leveraged for automated extraction of complex features from large amounts of high dimensional unsupervised data produced by IoT devices.

ML techniques can support self-organizing operations by learning and processing statistical information from the environment. These learning techniques are inherently distributed and are scalable for dense and large-scale IoT networks. Data-driven ML techniques are used to detect possible patterns and similarities in large data sets and can make predictions. ML is leveraged to develop algorithms that are capable of applying statistical methods on input data, performing analysis and predicting an output value without being explicitly programmed. Therefore, ML techniques require datasets to learn from and then the learned model is applied to the real data. However, the learned model may not generalize well to the entire range of features and properties of the data. In this regard, DL techniques have been employed to address some of the limitations of the ML techniques. Nonetheless, ML and DL algorithms have inherent limitations and also the development of these techniques brings along new challenges that must be addressed. In particular, data-driven ML and DL techniques require ``good quality" data which may not be available and also difficult to generate. For example, for a DL-based resource management method for channel and power allocation problem in a large-scale cellular IoT network, for a given network state (e.g. number of users, their association to the BS, channel gains to the respective BS), the optimal channel and power allocation data will be required to train the DNN. Generation of these data will be computationally too expensive even using techniques such as genetic algorithm(s). Therefore, the challenges related to data generation will need to be addressed to develop DL-based resource managewment schemes.   

Here we will focus on the challenges related to the development, (re)training, interoperability, cost, generalization, and the data-related issues of the ML models.

 \subsection{Challenges Associated With ML Models for IoT Networks}

\subsubsection{Development of ML models for IoT networks}
It is challenging to develop a suitable model to process data coming from a range of diverse IoT devices. Similarly, effective input data labeling is also a cumbersome task. 
For instance, in case of health-care applications, the training model must be pin-point accurate to derive the right features from the data and learn effectively, because false positives could have dire consequences in such models. Reliability of the ML models will be essential for mission-critical IoT applications which will be vulnerable to the anomalies (e.g. misclassifications) created because of ML or DL algorithms. Additionally, RL suffers from instability and divergence when all possible action-reward pairs are not used in a given scenario \cite{Mufti}. In a nutshell, depending on the use-case, the underlying IoT technology and the characteristics unique to that use-case, ML models must be efficiently crafted to meet the requirements of the use-cases. For the deployment of ML and DL models in the resource-constrained IoT devices, it is essential to reduce the processing and storage overhead \cite{Shuochao}. 

Again, engineering of the ML models (e.g. optimizing the {\em hyperparameters} in a DNN such as the number of hidden layers, number of neurons per layer, optimizing the learning rate in a DRL model) are challenging. 



\subsubsection{Lack of multitasking capabilities}
DL models are highly specialized and trained in a specific application domain. These models are only effective to solve one specific type of the problem for which it is trained. Whenever there is a change in the definition, state, or the nature of the problem (even if the new problem is very similar to the older one), re-assessment and retraining of the model is essential. Restructuring the architecture of the entire model is required when multitasking is done by the DL algorithms. For real-time IoT applications, it is difficult for the IoT devices to change and retain the DL models with frequently changing input information in a timely manner. Therefore, DL models capable of multitasking should be further investigated.

 \subsubsection{Retraining of ML and DL models} 
 When ML algorithms are applied to IoT applications and resource management in IoT networks, the models need to be updated according to the fresh collected data. This phenomenon gives rise to many important questions such as when and how ML models should be retrained with the changing environment and how to ensure that an appropriate amount of data is collected.

\subsubsection{Interpretability of ML and DL models}
When applying ML models to make informed decisions in IoT networks, in addition to accurate predictions, it is also helpful for the domain experts to gain insights into the decision making process of the models. In other words, interpretability of the model is very important to generate deep insights into the ML model. Powerful models such as DL and NNs are difficult to understand and interpret. Further research is required to build smart models which are not only highly accurate but also interpretable. The need for explainability of the ML models drives the need for simple models such as decision trees. It will be of fundamental importance to strengthen the existing theoretical foundations of DL with repeated experimental data. In this way, the performances of DL models can be quantified based on certain parameters such as computational complexity, learning efficiency, and accuracy.


\subsubsection{Cost of training of ML and DL models} 
Sufficient amount of data is required to train ML models and a considerable cost is associated with training these models. This training cost becomes more pressing, specifically in real-time IoT applications. Furthermore, the ML models are needed to be up-to-date in order to have accurate predictions over longer duration of time. Although model retraining can be done manually in the same way as the model was trained, it is a time consuming process. Another method is continuous learning in which an automated system is developed that continuously evaluates and retrains the existing models. Both of these methods require a well architected system model that should take into account not only the initial training procedure of a model but also a plan for keeping the models up-to-date.

\subsubsection{Generalization of ML models} 
Generalization is defined as the ability of an ML model to adapt to new and unseen data. The data is drawn from the same distribution used for training the model. Accuracy of the model is determined by how well it can be generalized to data points outside the training data sets. The generalization error is the measure of accuracy and the validity of an ML model. The propagation channels in a wireless IoT network can have varying and diverse physical characteristics. Therefore, although trained offline, the ML/DL models for resource management in IoT networks should be generalized to various channel conditions and must have dynamic adaptive capabilities. More research and implementation efforts are required to build practical models for real system with real user workloads. 

\subsection{Data Challenges in ML and DL-Enabled IoT Networks}

\subsubsection{Scarcity of datasets}
Learning algorithms entirely depend on the amount and quality of data to build models. Often, these datasets for IoT networks are scarce. Also, the availability of such datasets is subject to laws through which the data should be collected, the environment, and the consent of the data subject. Generation of synthetic datasets that could train models is also challenging. Another important issue is the uniformity of the datasets that could be used in cross-platform environments. 

 \subsubsection{Heterogeneity of IoT data}
Due to the massive and heterogeneous nature of IoT networks, the data generated by such networks has multiple dimensions and  ML models must be developed to extract useful information and features from the data. In this regard, pre-treatment of the data (pre-processing, cleansing, and ordering) and fusion of multi-source data will play an important role to make it ready for the respective ML model. 

\subsubsection{Collection of data}
 
The data generated in IoT networks can be used for a variety of different purposes depending on the application domain. For instance, some data maybe privacy-critical where it must be anonymized before using it for training. The example of such data is the data generated by smart-home environment and by the sensors on the human body. These data must be sent to central server(s) for further processing. Therefore, the context is also essential in the collection of data from different domains. Also, the ML and DL algorithms need to cope with anomalies in the data (e.g. data imbalance due to class skewness and class overlapping) in an intelligent manner. 

\subsubsection{Avalanche effect of ML and DL}

The avalanche effect is a desired characteristic of the cryptographic algorithms where a small change in the input causes huge change in the output. However, this effect is the default property of the ML and DL algorithms which also makes them vulnerable to adversarial attacks. In the context of IoT, this property of ML and DL mechanism will cause dire consequences to the applications. Therefore, the integrity of the data must be pre-checked before giving the data as input to the ML and DL algorithms. Recent perturbation attacks on NN \cite{Su2017} advocate that concrete steps are required to guarantee the integrity of input data to the ML and DL algorithms. One possible solution would be to take the context of the data into account and incorporate an effective access control mechanism to the input data.

\section{Conclusion}
\label{sec:conclusion}

The fundamental notion of connectivity among different smart devices with processing, storage, and communication capabilities, collectively constitute an IoT network. IoT has the potential to revolutionize many aspects of our lives ranging from our life-style, business, environment, infrastructure and so on. The heterogeneity introduced by the IoT is humongous from many aspects such as device types, network types, and data types. In the wake of such massive heterogeneous environment, IoT networks need scalable, adaptive, efficient, and fair network management mechanisms. Among other aspects of the network management, resource management has a pivotal role in the successful realization of massively heterogeneous IoT networks. In essence, IoT networks do not only face resource allocation challenges but also require contextual and situational custom resource management decisions. Unlike the traditional resource management methods including optimization and heuristics-based methods, game theoretical and cooperative approaches, the ML and DL models can derive actions from the run-time context information and can re-tune and re-train themselves with changes in the environment. ML and DL techniques are promising for automatic resource management and decision making process, specifically for large-scale, complex, distributed and dynamically changing IoT application environment. 

In this article, we have discussed the limitations of the traditional resource management techniques for wireless IoT networks. We have carried out a comprehensive survey of the ML- and DL-based resource management techniques in the IoT networks with enabling technologies such as HetNets, D2D, MIMO/massive MIMO, and NOMA. Furthermore, we have also covered different aspects of the resource management such as scheduling and duty cycling, resource allocation (in static and mobile IoT networks), clustering and data aggregation, resource discovery through spectrum sensing and cell selection, traffic classification and prediction, power allocation, and interference management. To this end, we have identified the future research challenges in applying ML and DL in IoT resource management. Applications of machine learning techniques, specifically deep reinforcement learning techniques, to solve complex radio resource management problems in emerging IoT networks is a fertile area of research.  For these networks, we will need to carefully craft the solutions according to the challenges such as model uncertainty, interpretability of the models, cost of model training, and generalization from test workloads to real application user workloads.


\bibliographystyle{IEEEtran}
\bibliography{References}
\end{document}